\documentclass[usenatbib]{mn2e}

\usepackage{amsmath}
\usepackage{natbib}
\usepackage{epsfig}
\usepackage{txfonts}
\usepackage{subfigure}
\usepackage{color}

\newcommand{\id}{{\,\rm d}}

\newcommand{\beq}{\begin{equation}}   %

\newcommand{\eeq}{\end{equation}}   %

\newcommand{\beqa}{\begin{eqnarray}}   %

\newcommand{\eeqa}{\end{eqnarray}}   %

\newcommand{\beal}{\begin{align}}

\newcommand{\enal}{\end{align}}

\newcommand{\bspl}{\begin{split}}

\newcommand{\espl}{\end{split}}

\newcommand{\bsub}{\begin{subequations}}

\newcommand{\esub}{\end{subequations}}

\newcommand{\bmulti}{\begin{multline}}   %

\newcommand{\beqm}{\begin{mathletters}}   %

\newcommand{\eeqm}{\end{mathletters}}   %

\newcommand{\Te}{T_{\rm e}}

\title[Aberrating the CMB sky]
{Aberrating the CMB sky: fast and accurate computation of the aberration kernel}

\author[Chluba]{J.~Chluba$^{1}$\thanks{E-mail: jchluba@cita.utoronto.ca}
  \\
$^{1}$ Canadian Institute for Theoretical Astrophysics, 60 St. George Street,
Toronto, ON M5S 3H8, Canada
}

\begin{document}

\date{Accepted 2011 April 18. Received 2011 March 31; in original form 2011 February 08}

\maketitle

\begin{abstract}
It is well known that our motion with respect to the cosmic microwave background (CMB) rest frame introduces a large dipolar CMB anisotropy, with an amplitude $\propto\beta=v/c\sim 10^{-3}$. {In} addition it should lead to a small {breaking of statistical isotropy} which becomes most notable at higher multipoles.
In principle this could be used to determine our velocity with respect to the CMB rest frame using high angular resolution data from {\it Planck}, without directly relying on the amplitude and direction of the CMB dipole, {allowing us to} constrain cosmological models in which the cosmic dipole arises partly from large-scale isocurvature perturbations instead of being fully motion-induced.
{Here we derive simple {recursion relations} that allow precise computation of the motion-induced coupling between different spherical harmonic coefficients.}
{Although the lowest order approximations for the coupling kernel can be deficient by factors of $2-5$ at multipoles $l\sim 1000-3000$, using our results for the aberration kernel we explicitly confirm that for a statistical detection of the aberration effect only first order terms in $\beta$ matter.}
{However, the expressions given here are not restricted to $\beta\sim 10^{-3}$, but can be used at much higher velocities. We} demonstrate the robustness of these formulae, illustrating the dependence of the kernel on $\beta$, as well as the spherical harmonic indices $l$ and $m$.

\end{abstract}


\section{Introduction}
The large dipolar temperature anisotropy of the cosmic microwave background (CMB) is usually interpreted as a consequence of our motion with respect to the CMB rest frame, implying that the Solar
System is moving with a velocity of $\beta=v/c=1.241\times 10^{-3}$ in the direction $(l, b)=(264.14^\circ\pm 0.15^\circ, 48.26^\circ\pm 0.15^\circ)$ \citep{Smoot1977,Fixsen1996, Scott2010}.
However, in addition our motion should lead to a small {\it breaking of statistical isotropy}, which is most notable at high multipoles $l$ or equivalently small angular scales, leading to the { coupling} of neighbouring spherical harmonic coefficients as a consequence of the { aberration} and {boosting} effect \citep{Challinor2002}.
In the direction of our motion, the temperature anisotropies are {beamed} towards each other, while in the opposite direction they are {magnified} (see Fig.~\ref{fig:illustrate_abberation} for illustration). 
This should lead to a tiny power asymmetry on the CMB sky \citep{Burles2006}, and introduces {correlations} between neighbouring spherical harmonic coefficients.

It was recently argued \citep{Kosowsky2010, Amendola2010} that the latter effect could be used to determine our velocity vector with respect to the CMB rest frame using high angular resolution data from the {\it Planck} satellite, without directly relying on the amplitude and direction of the CMB dipole.
This is because the CMB provides {\it both} the most cosmologically distant and the most statistically isotropic `marker' on the sky, and hence can be used to {search} for the small aberration effect.
In principle this should allow us to constrain cosmological models in which the cosmic dipole arises partly from large-scale isocurvature perturbations  \citep[e.g., see][and references therein]{Zibin2008} instead of being fully motion-induced.
Alternatively, one could use the small motion-induced asymmetry in the SZ cluster \citep{Zeldovich1969, Sunyaev1980} number counts to independently measure our velocity with respect to the CMB and place constraints on the primordial dipole. However, the required number of SZ clusters is rather large, which makes this endeavour difficult \citep{Chluba2005}.
{Also in the future, the cosmological 21cm signal from the reionization epoch could provide another opportunity to search for the aberration effect.}

To account for the effect of our motion on the CMB anisotropies {one has to} compute the amount of mixing between neighbouring spherical harmonic coefficients\footnote{{Alternatively, one can directly work in real space, however, here we follow the example of earlier works \citep{Challinor2002, Kosowsky2010, Amendola2010} on this problem.}}, $a_{lm}$. 
{Here we restrict ourselves to the CMB temperature anisotropies, however, it should be possible to extend our method to the case of polarization.}
It was shown earlier that {in} lowest order {of $\beta$ our motion} leads to a coupling\footnote{This assumes that the $z$-axis is aligned with the velocity vector.} of $a_{lm}$ with $a_{l\pm1m}$.
However, {it is difficult} to compute {the {\it aberration kernel\footnote{{At high $l$ the aberration effect dominates over the Doppler term. However, the aberration kernel as defined here includes the contributions from the Doppler term, which for the CMB statistics matters \citep{Amendola2010}.}}} which describes this coupling}, since the associated integrals are highly oscillatory, making numerical quadrature very demanding and time-consuming, even for the lowest order coupling terms, i.e. {between} $l\leftrightarrow l\pm 1$.
One way around this problem is to use series expansions of the corresponding integrals in orders of $\beta\ll 1$. Analytic expression accurate up to $\mathcal{O}(\beta^2)$ were obtained earlier, however, at large $l$ these expressions {converge slowly once $l\,\beta\gtrsim 1$ \citep{Challinor2002}}. 
Furthermore, in {recent discussions} of the aberration effect only the lowest order expressions, i.e. $\mathcal{O}(\beta)$, were applied. As we show here, these can be deficient by factors of $2-5$ at $l\sim 1000-3000$, and it becomes important to include higher order terms {when computing the coupling kernel}.
{However, for the statistical properties of the CMB {\it only} the first order terms in $\beta$ really matter, as we explicitly confirm here using our results for the kernel (see Sect.~\ref{sec:confirm}).}

For this purpose, we derive general recursion relations that allow accounting for terms up to high orders in $\beta$. 
Rather than focusing {on} $\beta\sim 10^{-3}$, we discuss the kernel for more general cases, showing that our method is both very fast and very robust, even for much higher velocities. 
We find that for precise computation of the aberration kernel at multipoles $l\gtrsim 1000-3000$ in the case of our motion with respect to the CMB rest frame terms up to {high orders in $\beta$ (e.g., $\mathcal{O}(\beta^{|l-l'|}\,\beta^{8})$ at $l\sim 3000$)} are required. 
We also illustrate the $l$, $m$ and $\beta$ dependence of the aberration kernel.

\section{The effect of (our) motion on the CMB sky}
In this section we briefly recap the key formulae to take the effect of motion on the CMB temperature anisotropies into account. Our formulation most closely reassembles the one in \citet{Kosowsky2010}, however, here we do not use a first order series expansion of the problem, but give general recursion formulae that in principle allow us to compute the coupling of different modes to machine precision for practically any value of $\beta$ (see Appendix~\ref{app:aberration_kernel}).
%

\subsection{Basic formulae}
In the CMB rest frame, $\mathcal{S}$, the energy spectrum of the CMB blackbody is given by
$I_\nu(\theta, \phi)\equiv B_\nu(T)$,
where $B_\nu$ is the blackbody spectrum of a given temperature at frequency $\nu$, and $T\equiv T(\theta, \phi)$ describes the CMB temperature in different directions of the sky.
Furthermore, one can write 
\bsub
\beal
T(\theta, \phi)&= T_0 [1+\Delta(\theta, \phi)],
\\
\Delta(\theta, \phi)&=\sum_{l=1}^{\infty} \sum_{m=-l}^{l} a_{lm} Y_{lm}(\theta, \phi),
\end{align}
\esub
using a spherical harmonic expansion of the temperature field. Here $T_0$ is the value of the CMB monopole in $\mathcal{S}$ and $\Delta$ describes the {primordial} CMB temperature anisotropies, {including a possible dipole asymmetry created, e.g., by large-scale isocurvature perturbations}.

To transform $I_\nu(\theta, \phi)$ into the moving frame\footnote{Henceforth primed quantities denote variables in the moving frame.} $\mathcal{S}'$ we use the Lorentz-invariance of the photon occupation number to obtain
\bsub
\beal
I'_{\nu'} (\theta', \phi') &=\frac{2h\nu'^3}{c^2} \frac{1}{e^{h\nu/kT}-1} 
\approx B_{\nu'}(T'_0)\left[1+ \mathcal{G}(x'_0)\Delta'(\theta', \phi') \right],
\end{align}
\esub
where $\mathcal{G}(x)=x\,e^x/[e^x-1]$ and $x'_0=h\nu'/kT'_0$. Here $T'_0$ defines the {apparent} monopole temperature\footnote{Although for our Solar System it is clear that $T'_0\approx T_0$, one can derive more general expressions that account for the {leakage} of power from the higher multipoles {into} the monopole term.} in $\mathcal{S}'$, and $\nu'$ denotes the frequency at which the measurement is performed. 
In addition, we have assumed that $\Delta' \ll 1$. However higher order corrections could in principle be taken into account, in next order leading to a $y$-type spectral distortion when comparing with a blackbody of temperature $T_0'$ \citep{Chluba2004}.

Rotating all $z$-axes parallel to the velocity vector of the moving frame one simply has $\phi\equiv \phi'$ and $\mu\equiv [\mu'+\beta]/[1+\beta\,\mu']$, where $\mu=\cos(\theta)$ and $\mu'=\cos(\theta')$.
The expression for {$T'$ and $\Delta'$ then read}
\bsub
\beal
T'(\theta', \phi')&= T'_0 [1+\Delta'(\theta', \phi')],
\\[1mm]
\Delta'(\theta', \phi') &= \frac{T_0}{T'_0}\, \frac{1+\Delta(\theta, \phi')}{ \gamma[1+\beta \mu'] } - 1,
\end{align}
\esub
{where we used} $\nu/T=\nu' \gamma [1+\beta \mu']/T \equiv \nu'/T'$. In addition, $\theta$ should be expressed as functions of $\theta'$ using the relations from above. 
To simplify matters further we write
\bsub
\label{eq:Delta_cont}
\beal
\Delta'(\mu', \phi') &= \Delta'_0(\mu', \phi') + \Delta'_{\rm an}(\mu', \phi'),
\\
\label{eq:Delta_cont_b}
\Delta'_0(\mu', \phi')&=  \frac{T_0}{T'_0}\, \frac{1}{ \gamma[1+\beta \mu']} - 1,
\\
\label{eq:Delta_cont_c}
 \Delta'_{\rm an}(\mu', \phi') &= \frac{T_0}{T'_0}\, \frac{\Delta(\mu, \phi)}{ \gamma[1+\beta \mu']}.
\end{align}
\esub
Here $\Delta'_0(\mu', \phi')$ is the temperature anisotropy in the moving frame that is arising from the {CMB rest frame monopole term, $\propto T_0$,} alone. Due to {Lorentz-boosting} it results in a motion-induced dipole, quadrupole, octupole, and higher order multipoles, all with increasing order of $\beta$. Since $\beta \ll 1$, for our Solar System one usually can stop after the motion-induced quadrupole. 
The contribution $ \Delta'_{\rm an}(\mu', \phi') $ arises from the primordial CMB anisotropies. Here both {boosting} and {aberration} terms are contributing, {with the aberration terms dominating at small scales}.

One can now perform a spherical harmonic expansion of the temperature field, $T'(\theta', \phi')$, to obtain the spherical harmonic coefficients $a'_{lm}(\beta)$ that describe the CMB sky inside $\mathcal{S}'$.
This yields
\bsub
\beal
\label{eq:A_an_def_a}
\Delta'(\theta', \phi')&=\sum_{l=0}^{\infty} \sum_{m=-l}^{l} a'_{lm}(\beta) \, Y_{lm}(\theta', \phi') - 1
\\
\label{eq:A_an_def_b}
a'_{lm}(\beta)&=\int Y^\ast_{lm}(\mu', \phi')\, \Delta'(\mu', \phi')  \id \Omega' 
\nonumber\\
&= \sum_{l'=0}^{\infty} \sum_{m'=-l'}^{l'} \mathcal{K}^{l'm'}_{lm}(\beta)\,a_{l'm'} 
   \equiv \sum_{l'=0}^{\infty} \mathcal{K}^{l'm}_{lm}(\beta)\,a_{l'm}.
\end{align}
\esub
The {\it aberration kernel} $\mathcal{K}^{l'm'}_{lm}(\beta)$ is defined in Appendix~\ref{app:aberration_kernel}. 
For the last equality, we have used the fact that in the chosen frame the kernel only mixes the different $l$ and $l'$ for fixed $m$ (see Appendix~\ref{app:aberration_kernel}). 
{The general case can be obtained by rotating the kernel.}
{Note that $a_{00}=\sqrt{4\pi}$ by construction.}

Equation~\eqref{eq:A_an_def_b} shows that the problem is fully determined {once} all elements of the aberration kernel {are known}.
In Appendix~\ref{app:aberration_kernel} we explain how to obtain these using simple recursion relations.

\begin{figure}
\centering
\includegraphics[width=\columnwidth]{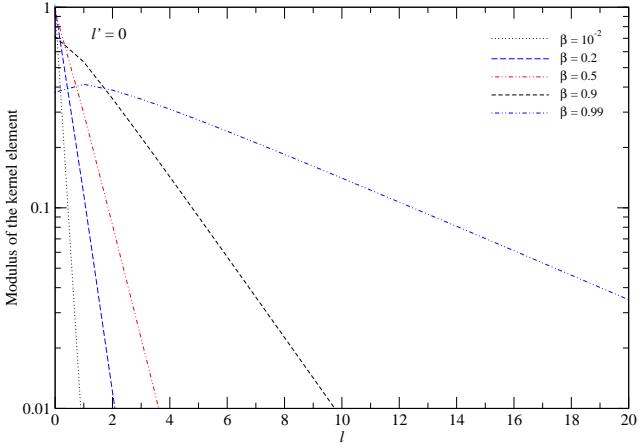}
\caption{Modulus of the aberration kernel $\mathcal{K}^{00}_{l0}(\beta)$ for different {$l$} and increasing value of $\beta$.
In the computation $k=256$ was used for $\beta=0.99$ to reach convergence.}
\label{fig:Kernel_0_beta_depend}
\end{figure}
\subsection{CMB monopole term}
To understand the effect of the motion of our Solar System on the CMB anisotropies we start by writing down the motion-induced signal arising from the CMB monopole alone.
Although this case can be easily treated using a series expansion in terms of $\beta$, it illustrates the procedure in the more general case.

To give the solution to the problem we need the kernel elements $\mathcal{K}^{00}_{l0}(\beta)$.
From Eq.~\eqref{eq:redef_H} one directly has
\beal
\label{eq:K_00}
\mathcal{K}^{00}_{00} (\beta)
&=\frac{T_0}{T'_0\gamma}\,\sum_{k=0}^\infty  \frac{\beta^{2k}}{2k+1}
   =\frac{T_0}{T'_0} \,\frac{{\rm arctanh(\beta)}}{\beta\gamma} \approx \frac{T_0}{T'_0}[1-\beta^2/6],
\end{align}
with ${}^{(0)} \tilde{\kappa}^{0\rightarrow 0}_{2k}=1/[2k+1]$. This result can also be easily verified by analytic integration.
To obtain the elements $\mathcal{K}^{00}_{l0}(\beta)$ for $l>0$ we can use the recursion formula Eq.~\eqref{eq:rec_kappa_00_l}. With the initial condition ${}^{(0)} \tilde{\kappa}^{0\rightarrow 0}_{0}=1$ it is straightforward to show that 
\bsub
\beal
\label{eq:kappa_00_k}
{}^{(0)} \tilde{\kappa}^{0\rightarrow l}_0&=\frac{g_l \,l! }{(2l+1)!!}
&{}^{(0)} \tilde{\kappa}^{0\rightarrow l}_2&=\frac{ g_l \,(l+2)!}{2 (2l+3)!!}
\\
{}^{(0)} \tilde{\kappa}^{0\rightarrow l}_4&=\frac{ g_l \,(l+4)!}{8\,(2l+5)!!}
&{}^{(0)} \tilde{\kappa}^{0\rightarrow l}_6&=\frac{ g_l \,(l+6)!}{48\,(2l+7)!!},
\end{align}
\esub
with $g_l=\sqrt{2l+1}$.
With these coefficients one can determine all $\mathcal{K}^{00}_{l0}(\beta)$ up to $12^{\rm th}$ order in $\beta$, while higher orders could be easily obtained with Eq.~\eqref{eq:rec_kappa_00_l}.  
%
%
Furthermore, according to Eq.~\eqref{eq:prop_b} one directly has the kernel elements $\mathcal{K}^{l0}_{00}(\beta)=(-1)^l\mathcal{K}^{00}_{l0}(\beta)$, which allow us to compute the leakage of power from the higher multipoles to the apparent CMB monopole.
Putting everything together we find
\beal
T'_0 &=  T_0 \frac{{\rm arctanh}(\beta)}{\beta\gamma}
+ T'_0 \,\sum_{l=1}^{\infty} (-1)^l\, {\mathcal{K}^{00}_{l0}(\beta)}\, \frac{a_{l0}}{\sqrt{4\pi}} 
\nonumber \\&
\approx T_0\left[ 1- \frac{\beta^2}{6} 
+ \frac{\beta}{\sqrt{3}} \,\frac{a_{10}}{\sqrt{4\pi}} 
+  \frac{2\beta^2}{3\sqrt{5}} \,\frac{a_{20}}{\sqrt{4\pi}} \right]
\end{align}
for the value of the apparent CMB monopole in $\mathcal{S}'$.
Clearly higher order terms are very small in the case of our Solar System, and as Fig.~\ref{fig:Kernel_0_beta_depend} shows, even for $\beta\sim 0.9$ only about half of the intrinsic dipole is expected to leak into the monopole term.
Similarly, we find
\beal
1\!+ \!\Delta'_0(\theta', \phi')&=  \frac{T_0}{T'_0} \frac{{\rm arctanh}(\beta)}{\beta \gamma}
+ \sum_{l=1}^{\infty}  \sqrt{4\pi}\;\mathcal{K}^{00}_{l0}(\beta)\, Y_{l0}(\theta', \phi')
\nonumber\\
&\approx \!1
- \sqrt{4\pi}\frac{\beta}{\sqrt{3}} Y_{10}(\theta', \phi')
+\sqrt{4\pi}\frac{2\beta^2}{3\sqrt{5}}  Y_{20}(\theta', \phi'),
\end{align}
where the sum accounts for the motion-induced anisotropies arising from the CMB monopole only (i.e. leakage of power from $l'=0$ to $l>0$).
{Here we approximated $T_0\approx T'_0$.}
As is well known, to leading order only the motion-induced dipole really matters in the case of our Solar System.
However, the recursions given here also allow us to compute the effect for large $\beta$.
For example, a fast moving electron inside the hot gas of a galaxy cluster (with $k\Te\sim 5-15\, \rm keV$) can have velocities $\beta \sim 0.1$. Because of the leakage of power from the monopole term to higher multipoles the electron `sees' a reduced CMB monopole, while dipole, quadrupole and higher multipoles are increased accordingly. In the rest frame of the moving electron the scattering process is given by {Thomson} scattering, implying that only the value of the rest frame monopole and quadrupole matters. 
This change of the CMB monopole and quadrupole in the rest frame of the scattering electron is one {reason} for {\it relativistic corrections} \citep{Challinor1998, Itoh1998, Sazonov2000, Chluba2005} to the SZ signals from galaxy clusters \citep{Zeldovich1969, Sunyaev1980}.

Note that the change in the apparent monopole temperature also means that, as a matter of principle, one should correct the observed monopole to the CMB rest frame value when calculating the CMB power spectra. However, in practice this correction will be below the cosmic variance level of the monopole \citep{Zibin2008}, and hence is  negligible (D. Scott, private communication).

\begin{figure}
\centering
\includegraphics[width=\columnwidth]{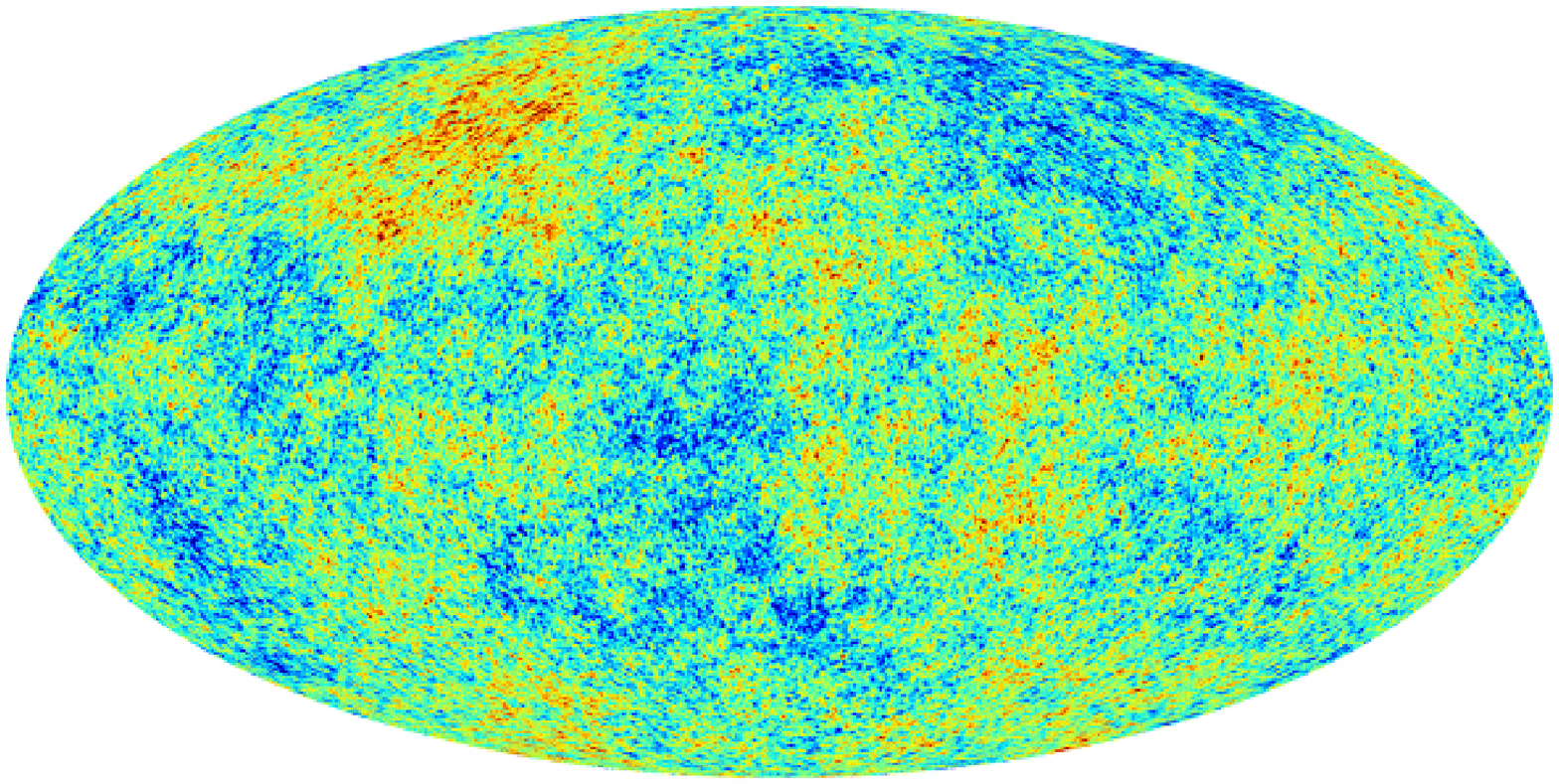}
\\[2mm]
\includegraphics[width=\columnwidth]{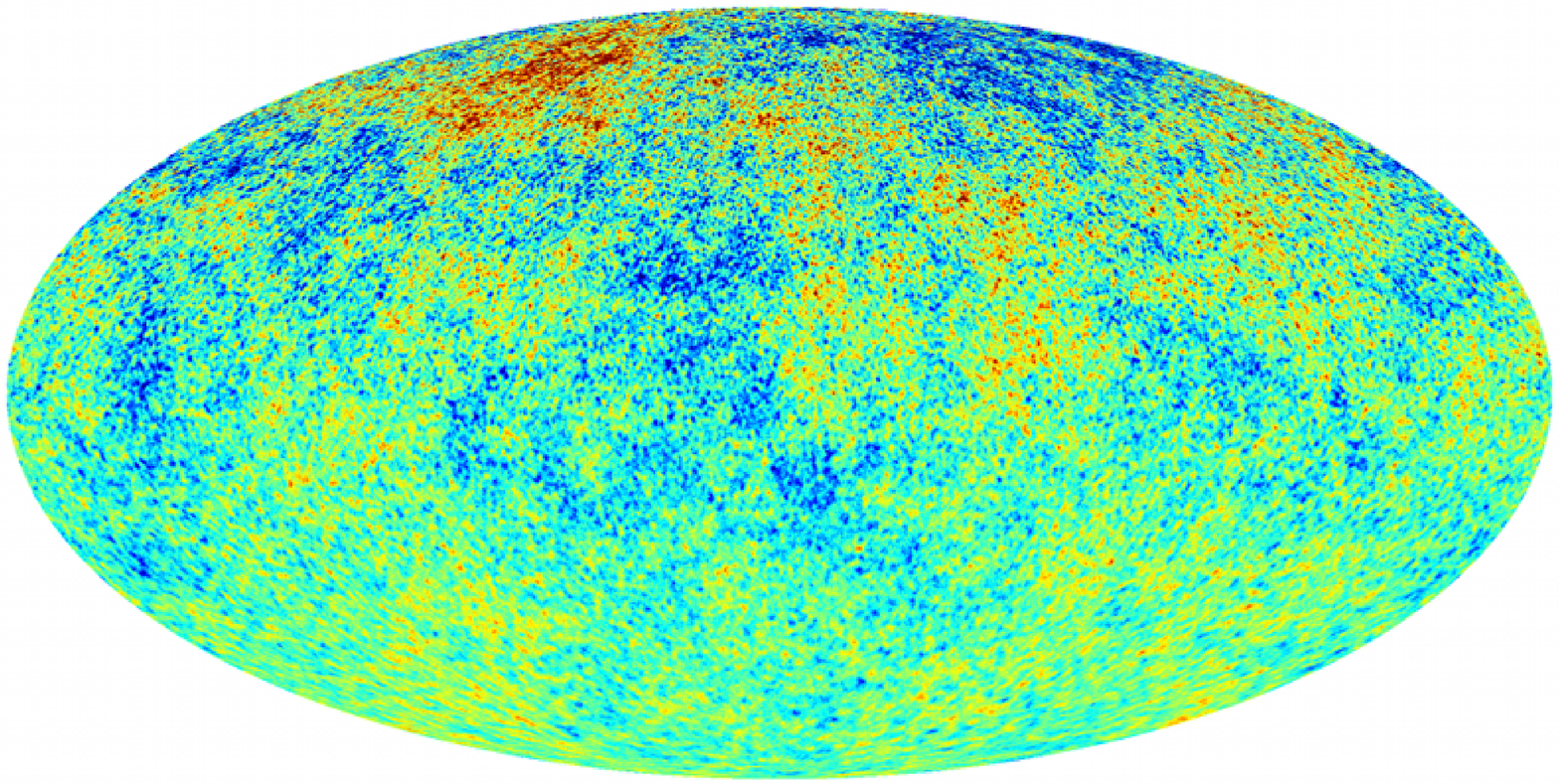}
\\[2mm]
\includegraphics[width=\columnwidth]{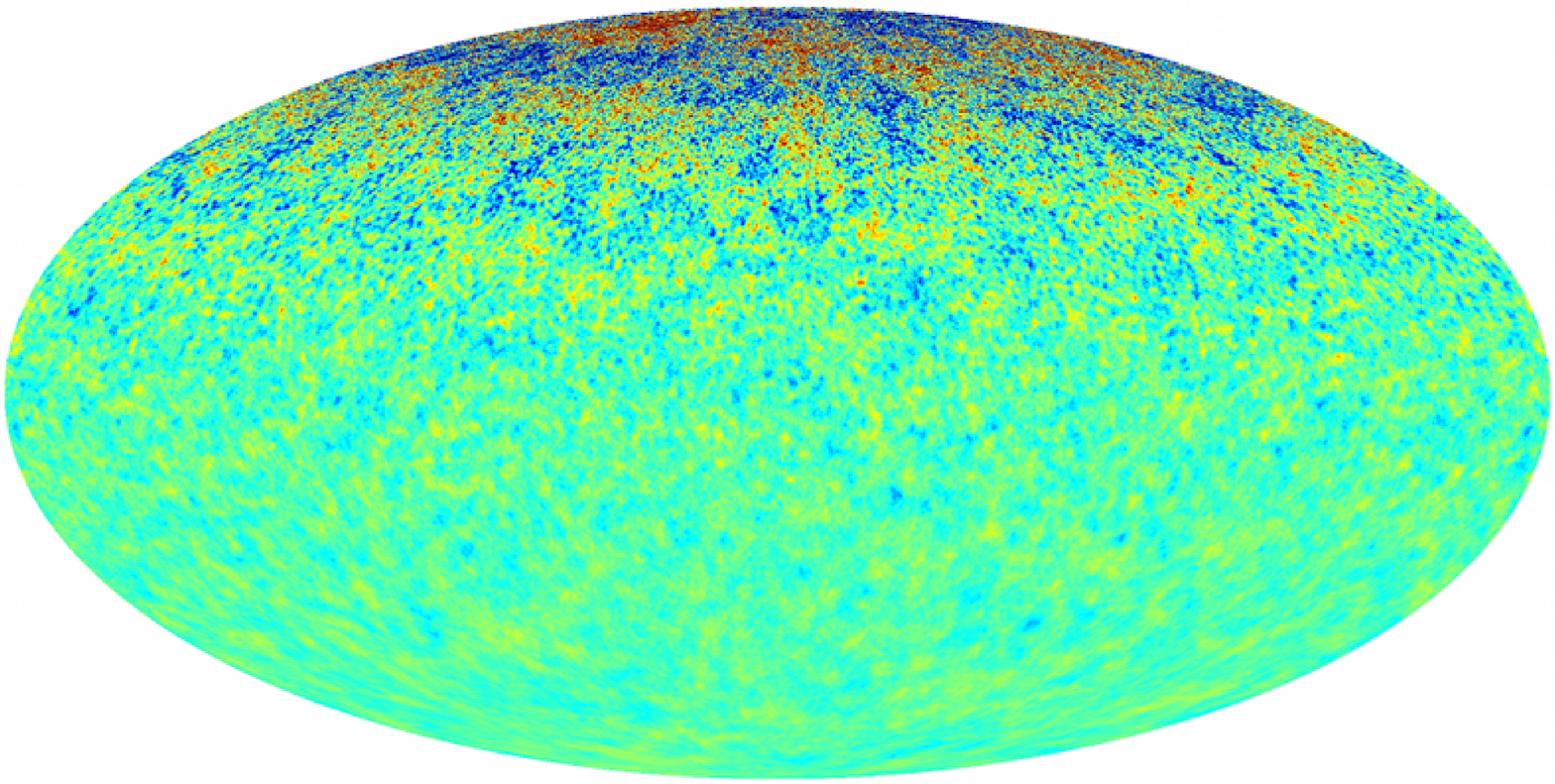}
\caption{Illustration of the effect of motion on the CMB temperature anisotropies for high velocities along the $z$-direction towards the north pole. The upper panel shows the CMB in its rest frame, the middle panel is for $\beta=0.5$, and the lower panel for $\beta=0.9$. In all cases the monopole and dipole terms were not included. The maps were created using a modified version of {\sc Healpix} \citep{Gorski2005}.}
\label{fig:illustrate_abberation}
\end{figure}
\subsection{Motion-induced terms for higher multipoles}
In this section we present several results for the aberration kernel, illustrating its dependence on $l$, $l'$, $m$ and $\beta$. These were obtained using the recursion relations given in Appendix~\ref{app:aberration_kernel}, however, we checked the precision of the results in several cases using explicit numerical integration, finding relative differences $\lesssim 10^{-10}$ when using a large number\footnote{Depending on the value for $\beta$ this could mean $k\sim 4$ for $l\lesssim 3000$, but also $k\sim 256$ for large $l$ and $\beta$.} of terms in the recursions. 

\subsubsection{Illustration of the aberration and boosting effect}
To illustrate the effect of motion on the CMB temperature anisotropies, we modified {\sc Healpix} \citep{Gorski2005} to allow accounting for the {\it aberration} and {\it boosting} effects caused by the Lorentz transformation of the CMB sky into the moving frame.
The results are shown in Fig.~\ref{fig:illustrate_abberation}. We chose very high velocities here to clearly illustrate the effect.
In the direction of the motion (upwards) the anisotropies are {\it beamed} towards each other, while in the opposite direction the effect of aberration acts like a {\it magnifying glass}.
One can clearly see a asymmetry between the power in the southern and northern hemispheres.
Of course, for our Solar System $\beta\sim 10^{-3}$ and the effect is much smaller and can only be picked up by looking at the breaking of statistical isotropy. Also the power asymmetry will be much smaller.
{While the asymmetry is expected to be $\propto 2\beta$ when computing the difference between the power spectra on the two hemispheres, on the full sky it is} only of order $\beta^2$ \citep{Challinor2002, Burles2006}.

\begin{figure}
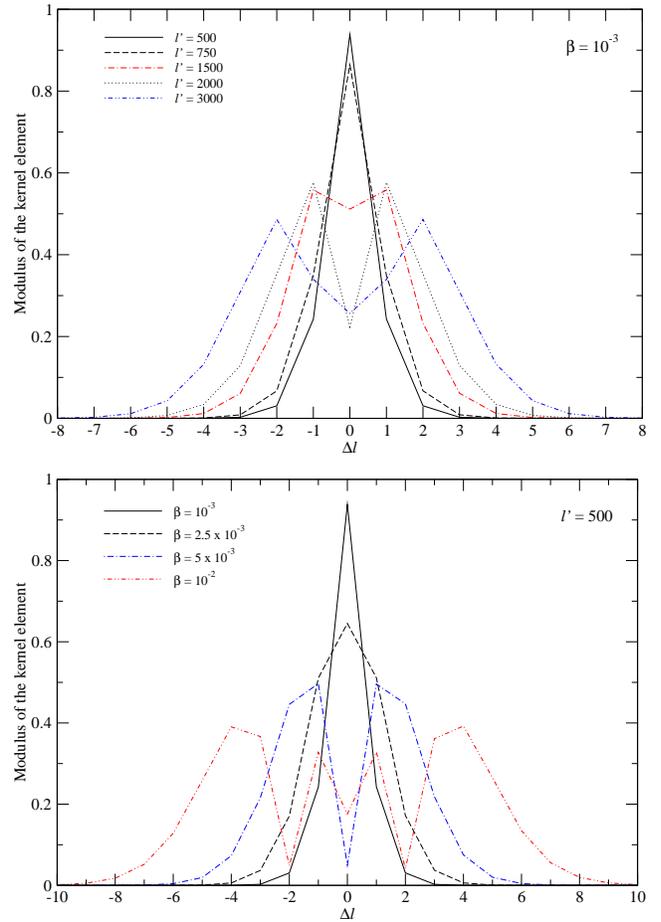

\centering
\includegraphics[width=\columnwidth]{./eps/K_beta_0.001.diff_l.eps}
\\[2mm]
\includegraphics[width=\columnwidth]{./eps/K_different_beta.eps}
\caption{Modulus of the aberration kernel $\mathcal{K}^{l'0}_{l0}(\beta)$ for different  $\Delta l=l-l'$. Upper panel: for different $l'$ and $\beta=10^{-3}$. Lower panel: for $l'=500$ and different values of $\beta$.}
\label{fig:Kernel_lp_beta_depend}
\end{figure}
\subsubsection{Dependence on $l'$ and $\beta$ for $m=0$}
In Fig.~\ref{fig:Kernel_lp_beta_depend} we show the behaviour of the kernel, $\mathcal{K}^{l'0}_{l0}(\beta)$, with different $l'$ and $\beta\sim 10^{-3}$ (upper panel) and for $l'=500$ but varying $\beta$ (lower panel).
For clarity we have plotted the modulus of the kernel to suppress its alternating behaviour.
{As was expected,}  the kernel becomes broader with increasing $l'$ and fixed value of $\beta$, and similarly, for fixed $l'$ but increasing $\beta$. Also the modulus of the kernel appears to be quasi-symmetric, however, this symmetry is not perfect, since there is a small but non-vanishing derivative of $\mathcal{K}^{l'0}_{l0}(\beta)$ with respect to $l'$, which makes $|\mathcal{K}^{l'0}_{l'+\Delta l \, 0}(\beta)|\neq |\mathcal{K}^{l'0}_{l'-\Delta l \, 0}(\beta)|$.
{For practical purposes this quasi-symmetry could be used to compress the kernel and minimize storage.}

Figure \ref{fig:Kernel_lp_beta_depend} also shows that even for $\beta=10^{-3}$ (i.e. close to the value of our own motion with respect to the CMB rest frame) at $l'\gtrsim 1000$ the coupling of $l'\rightarrow l'\pm2$ becomes important.
For $l'\sim 3000$ it is even stronger than the coupling $l'\rightarrow l'\pm1$, and also the $l'\rightarrow l'\pm3$ and $l'\rightarrow l'\pm4$ terms start to be significant.
As mentioned above, this demonstrates that for $l'\gtrsim 1/\beta$ higher order terms in the series become {important.}
%

\begin{figure}
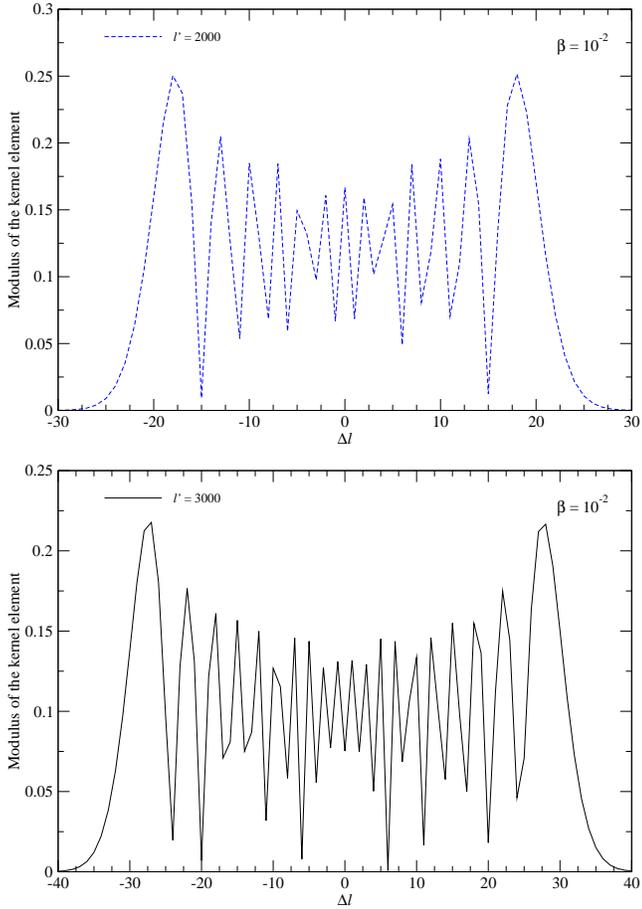

\centering
\includegraphics[width=\columnwidth]{./eps/K_m.lp_2000.b_0.01.eps}
\\[2mm]
\includegraphics[width=\columnwidth]{./eps/K_m.lp_3000.b_0.01.eps}
\caption{Modulus of the aberration kernel $\mathcal{K}^{l'0}_{l0}(\beta)$ for $\beta=10^{-2}$ and large $l'$. In the computations we used $k=64$ to reach convergence.}
\label{fig:Kernel_large_beta_lp}
\end{figure}
To push this point even further, in Fig.~\ref{fig:Kernel_large_beta_lp} we {show the aberration kernel} for $\beta=10^{-2}$. 
One can clearly see that the kernel becomes very wide, rendering a perturbative analytic expansion difficult.
In the recursions we used $k=64$, i.e. included terms $\xi=l'\beta$ up to $\xi^{128}$. The computation for {\it all} elements, $\mathcal{K}^{l'0}_{l0}(\beta)$, with $l'\leq l$ and $|\Delta l|\leq 100$ up to $l'=3000$, takes a few seconds using the recursion formulae, while brute force numerical integration just for $l'=3000$ takes hours with {\sc Mathematica}.
We also implemented an integration scheme based on Chebyshev {quadrature}, however, also in this case the computation takes too long for real applications.

\begin{figure}
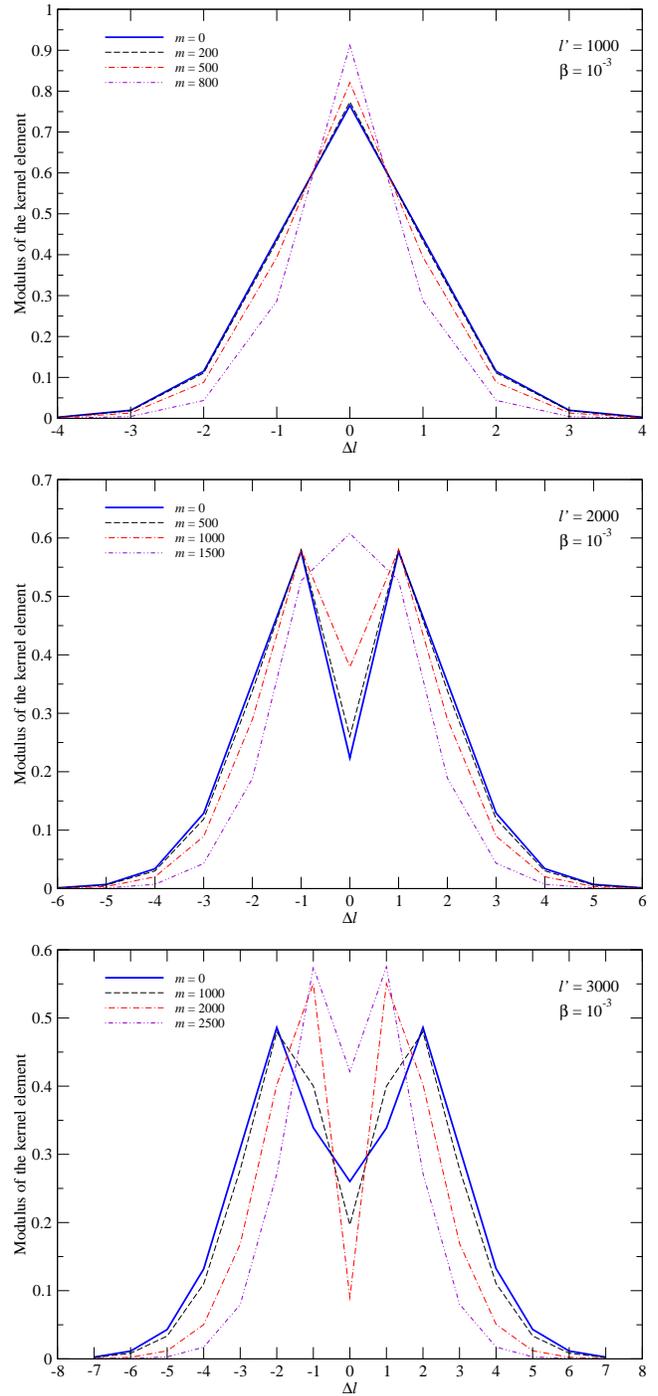

\centering
\includegraphics[width=\columnwidth]{./eps/K_m.lp_1000.b_0.001.m_depend.eps}
\\[2mm]
\includegraphics[width=\columnwidth]{./eps/K_m.lp_2000.b_0.001.m_depend.eps}
\\[2mm]
\includegraphics[width=\columnwidth]{./eps/K_m.lp_3000.b_0.001.m_depend.eps}
\caption{Modulus of the aberration kernel $\mathcal{K}^{l'm}_{lm}(\beta)$ for fixed $\beta=10^{-3}$ and different $l'$ and $m$. In the recursions we used $k=6$.}
\label{fig:Kernel_m_depend}
\end{figure}
\subsubsection{Dependence on $m$}
In Fig.~\ref{fig:Kernel_m_depend} we illustrate the $m$-dependence of the aberration kernel. For fixed $l'$, the kernel becomes narrower with increasing $m$. Also it is clear that the variation of the kernel with $m$ is rather slow. Apparently the important parameter is {$m/l$}.
This fact allows a strong compression of the kernel functions, e.g., for $l'=2000$ the kernel for $m\sim 0-500$ is practically not changing. Therefore one only has to store the coefficients ${}^{(m)}\! \kappa^{l' \rightarrow l}_{2k}$ for a fractions of the kernel elements, to obtain very precise results.
Numerical computations can benefit from this property, allowing memory consumption to be reduced.

\begin{figure}
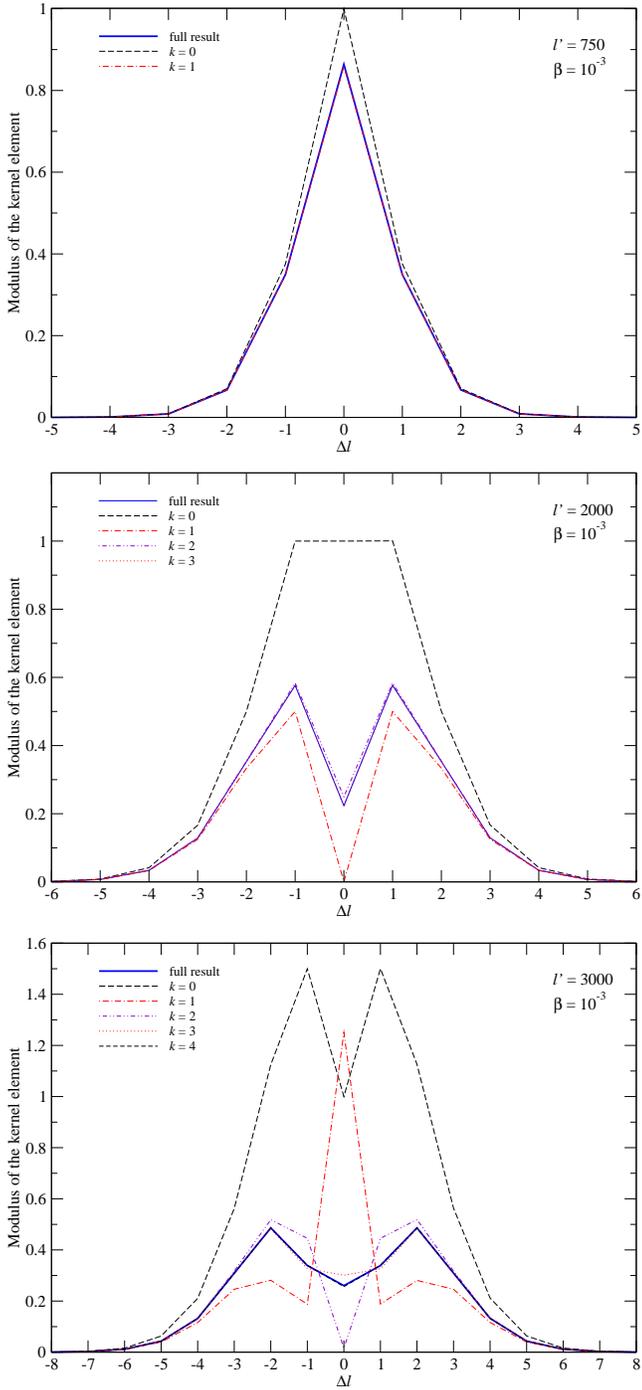

\centering
\includegraphics[width=\columnwidth]{./eps/K_beta_0.001.converge.lp_750.eps}
\\[2mm]
\includegraphics[width=\columnwidth]{./eps/K_beta_0.001.converge.lp_2000.eps}
\\[2mm]
\includegraphics[width=\columnwidth]{./eps/K_beta_0.001.converge.lp_3000.eps}
\caption{Modulus of the aberration kernel $\mathcal{K}^{l'0}_{l0}(\beta)$ for $\beta=10^{-3}$ and different $l'$. In this figure we illustrate the convergence to the series expansion, Eq.~\eqref{eq:redef_H}. For $l'\gtrsim 1000$ several correction terms have to be taken into account to obtain precise results.}
\label{fig:Kernel_convergence}
\end{figure}
\subsubsection{Convergence of the recursions}
We already mentioned several times that for large $l'$ and/or large $\beta$ many terms in the series have to be computed. This is because the coefficients ${}^{(m)}\! \tilde{\kappa}^{l'\rightarrow l}_{2k}$ strongly increase with $k$.
In Fig.~\ref{fig:Kernel_convergence} we show the convergence of the aberration kernel for $\beta=10^{-3}$ at representative values of $l'$. At $l'\sim 750$ the lowest order terms (i.e. $\mathcal{O}(\beta)$ for $l'\rightarrow l'\pm 1$; $\mathcal{O}(\beta^2)$ for $l'\rightarrow l'\pm 2$, etc.) already give rather good results for the strength of the mode coupling. We note that even for $l'\sim 750$ the kernel element for $l'\rightarrow l'\pm 2$ has amplitude $\sim 10\%$.

{
Going to higher values of $l'$ the kernel becomes broader, as explained above. 
For $l'\sim 2000-3000$ terms up to {$\mathcal{O}(\beta^{|l-l'|}\,\beta^{8})$} are important for accurate computations of the aberration kernel. The recursions given here easily allow such precision, while being sufficiently simple.
Furthermore, as our computations show, a lowest order expansion will not allow us to compute the precise value of the coupling kernel once $l'\gtrsim 1000$, as also mentioned earlier \citep{Challinor2002}.
We show this more quantitatively in Table~\ref{tab:kernel}, where we compare the results for the kernel obtained with the approximations given by \citet{Kosowsky2010} to our result for $k=6$. The case of \citet{Kosowsky2010} is equivalent to our case $k=0$, and we confirmed the values by direct comparison with their expressions.
As Table~\ref{tab:kernel} clearly shows, the lowest order expansion can be deficient by a factor of $2-5$.
}

\begin{table}
\centering
\caption{Representative kernel elements for $l'\rightarrow l'$ and $l'\rightarrow l'+1$. We compare the result obtained with the {approximations Eq.~(13) and (14) in} \citet{Kosowsky2010} with those from this work up to $k=6$. In some cases the lowest order expansions are deficient by a factor of a few. For $l'\gtrsim 2500$ even the sign changes for the kernel element $\mathcal{K}^{l'0}_{l'0}(\beta)$.}
\begin{tabular}{lccc}
\hline
$l'$ & $\Delta l$ & {$\mathcal{K}^{l'0}_{l0}(\beta)$ in $\mathcal{O}(\beta)$}  &  {$\mathcal{K}^{l'0}_{l0}(\beta)$ in $\mathcal{O}(\beta^7)$}   
\\
\hline 
1500 & 0 & 1 & 0.51155
\\
1500 & 1 & $-0.75050$ & $-0.55808$
\\
\hline 
2000 & 0 & 1 & 0.22360
\\
2000 & 1 & $-1.00050$ & $-0.57666$
\\
\hline 
2500 & 0 & 1& $-0.04863$
\\
2500 & 1 & $-1.25050$ & $-0.49685$
\\
\hline 
3000 & 0 & 1 & $-0.26021$
\\
3000 & 1 & $-1.50050$ & $-0.33869$
\\
\hline
\end{tabular}
\label{tab:kernel}
\end{table}

{
\section{Signal covariance matrix and measurements of $\beta$.}
\label{sec:confirm}
}
{Although the values of the kernel elements computed to first order in $\beta$ can be deficient by factors of $2-5$ (see previous Section), for a statistical detection of the aberration effect it is important how the signal covariance matrix is affected. 
In real space the terms arising because of aberration can be modelled as dipolar convergence \citep[e.g., see][]{Amendola2010}. In this case, analogy to the CMB lensing effect \citep{Lewis2006lens} implies that higher order terms in $\beta$ will not be important for a statistical detection of the aberration effect, as order by order in $\beta$ leading terms cancel.}

{
As pointed out by  \cite{Amendola2010} at high $l$ even the Doppler term matters. 
The Doppler term is not directly analogous to a lensing effect and hence the simple argument about its importance for the signal covariance matrix is not evident.
However, one can use $1/[1+\beta \mu']= \sum_{k=0}^\infty (-1)^k (\beta \mu')^k$ in the definition of the kernel, Eq.~\eqref{eq:Kernel_def_1}.
Repeatedly applying the recursion relations for the associated Legendre polynomials allows absorbing each of the extra factors of $\mu'$, order by order.
In this way it is clear that also the Doppler term essentially lead to a coupling of neighbouring modes that is similar to an aberration effect, however, in each order of $\beta$ it is about $l$ times smaller than the aberration effect.
Still in the covariance matrix it contributes at the same level \cite{Amendola2010}.
Again by analogy with the CMB lensing effect this implies that also higher order terms from the Doppler effect lead to small corrections in the covariance matrix, in agreement with earlier works on this problem \citep{Challinor2002}.
}

{With our results for the kernel elements we were able to confirm this statement using explicit computation of the covariance matrix elements $\left< a'^\ast_{lm} a'_{l+1\,m}\right>$. Although we explicitly use coupling terms up to $\Delta l'=10$ obtained with our recursions, the final covariance matrix element agrees with the simple first order result }
\beal
\label{eq:covar}
{
\left< a'^\ast_{lm} a'_{l+1\,m}\right>\approx -\beta (l+1) \sqrt{\frac{(l+1)^2-m^2}{4(l+1)^2-1}} \left[ C_{l+1} - C_{l} \right]
}
\end{align}
{where $C_l$ denotes the CMB temperature power spectrum. We found that all other off-diagonal covariance matrix elements with $l'>l+1$ for an ideal experiment do not affect the signal-to-noise ratio by more than $\sim 0.5\%-1\%$ for $l\lesssim 3000$, and hence can be neglected.}

\section{Conclusions}
We have obtained simple recursion relations that allow computing the elements of the aberration kernel in a very fast and accurate way.
The application of these recursions is not restricted to the small $\beta$ case, but in principle {enable precise} computations of the aberration effect for general $\beta$.
Here we illustrate the main properties of the aberration kernel for a wide range of parameters. 

{
Using the recursion relations we show that the lowest order expansions  for the couplings between spherical harmonic coefficients $l'\rightarrow l'$ and $l'\rightarrow l'\pm 1$ can be deficient by factors of $2-5$.
For $l'\sim 2000-3000$ terms up to $\mathcal{O}(\beta^{|l'-l|}\,\beta^8)$ are important for accurate computations of the aberration kernel. The recursions given here easily allow achieving such precision, while being sufficiently simple. }
{Albeit this large difference in the values of the kernel, for a statistical detection of the aberration effect the only the covariance matrix elements $\left< a'^\ast_{lm} a'_{l+1\,m}\right>$ really matter, and the first order expansion in $\beta$ provides a sufficient estimate.}

\section*{Acknowledgements}
{JC would like to thank the anonymous referee for pointing out weaknesses in our analysis 
regarding the importance of higher order coupling terms for a statistical detection of the aberration effect.
JC would also like to thank Martin Reinecke for helping to modify {\sc Healpix} in summer 2004.
Furthermore, he is very grateful for stimulating and encouraging discussions about this problem with Ue-Li Pen and Douglas Scott, and their detailed comments on the manuscript.
In addition, he would like to thank Duncan Hanson, Antony Lewis, Yin-Zhe Ma, Adam Moss, Geoff Vasil and James Zibin for useful discussions and comments.
Furthermore, he is very grateful for additional financial support from
the Beatrice~D.~Tremaine fellowship 2010.  
}

\bibliographystyle{mn2e}
\bibliography{Lit}

\begin{appendix}

\onecolumn

\section{Definition and computation of the aberration kernel}
\label{app:aberration_kernel}
The effect of our motion on the CMB temperature anisotropies can be fully characterized by the {\it aberration kernel}, which is defined as
\beal
\label{eq:Kernel_def_1}
\mathcal{K}^{l'm'}_{lm}(\beta)
&=
\frac{T_0}{T'_0 \gamma}
\int \frac{Y^\ast_{lm}(\mu', \phi')\,Y_{l'm'}(\mu, \phi')}{1+\beta \mu'}  \id \Omega',
\end{align}
where $\mu=\mu(\mu', \beta)=(\mu'+\beta)/(1+\beta\mu')$.
Using the definition for the spherical harmonic functions
\bsub
\label{eq:Norm_fact}
\beal
Y_{lm}(\mu, \phi) &= \mathcal{N}_{lm} \, P^m_l(\mu) \, e^{i\,m\,\phi},
\\
\mathcal{N}_{lm}&=\sqrt{\frac{(2l+1)}{4\pi}\,\frac{(l-m)!}{(l+m)!}},
\end{align}
\esub
where $P^m_l(\mu)$ are the associated Legendre polynomials, it is easy to show that
\bsub
\label{eq:Kernel_def_2}
\beal
\mathcal{K}^{l'm'}_{lm}(\beta)
&=\frac{2\pi\,T_0}{T'_0 \gamma}\,\delta_{mm'}\,\mathcal{N}_{lm}\,\mathcal{N}_{l'm}
\,\mathcal{H}^m_{l'\rightarrow l}(\beta),
\\[1mm]
\label{eq_def_Hlm}
\mathcal{H}^m_{l'\rightarrow l} (\beta)&=\int_{-1}^{1} \frac{P^m_l(\mu')\,P^m_{l'}(\mu)}{1+\beta \mu'}  \id \mu'.
\end{align}
\esub
The problem is now that the explicit computation of the integrals $\mathcal{H}^m_{l'\rightarrow l} (\beta)$ is a cumbersome task, since for large $l,l'$ and {$|m| \ll l, l'$} one is dealing with highly oscillatory functions. However, as we will show below, it is possible to obtain simple recursion relations that avoid this problem and allow precise and fast computation of $\mathcal{H}^m_{l'\rightarrow l} (\beta)$ to machine precision for practically any value of $\beta$.

\subsection{Simple properties of the integrals $\mathcal{H}^m_{l'\rightarrow l} (\beta)$}
\label{app:properties_kernel}
Knowing the properties of the associated Legendre polynomials, it is simple to deduce some useful properties of the integrals $\mathcal{H}^m_{l'\rightarrow l} (\beta)$, and hence the aberration kernel.
From $P^m_l(-x)=(-1)^{l+m}\,P^m_l(x)$ it directly follows that 
\beal
\label{eq:H_prop_a}
\mathcal{H}^m_{l'\rightarrow l} (-\beta)=(-1)^{l+l'}\,\mathcal{H}^m_{l'\rightarrow l} (\beta).
\end{align}
This implies that the {\it de-aberration} kernel $\mathcal{K}^{l'm'}_{lm}(-\beta)$ can be directly obtained from $\mathcal{K}^{l'm'}_{lm}(\beta)$.
Furthermore, with the identity $\id \mu' / [1+\beta\mu' ]= \id \mu/[1-\beta\mu] $ one has 
\beal
\label{eq:prop_b}
\mathcal{H}^m_{l'\rightarrow l} (\beta)&\equiv \int_{-1}^{1} \frac{P^m_l(\mu'(\mu,\beta))\,P^m_{l'}(\mu)}{1-\beta \mu}  \id \mu
\nonumber\\
&=\mathcal{H}^m_{l\rightarrow l'} (-\beta)
=(-1)^{l+l'}\mathcal{H}^m_{l\rightarrow l'} (\beta).
\end{align}
This relation reduces the number of independent kernel elements by a factor of two.
For example, knowing how the CMB monopole leaks into the higher multipoles, i.e. knowing $\mathcal{H}^0_{0\rightarrow l} (\beta)$, we conveniently have $\mathcal{H}^0_{l\rightarrow 0} (\beta)=(-1)^l\,\mathcal{H}_{l} (\beta)$ for all $l$.

Also, with $P^{-m}_{l}(x)=(-1)^m \frac{(l-m)!}{(l+m)!}\,P^{m}_{l}(x)$ we obtain 
\beal
\mathcal{H}^{-m}_{l\rightarrow l'} (\beta)&=\frac{(l-m)!}{(l+m)!}\frac{(l'-m)!}{(l'+m)!}\, \mathcal{H}^m_{l\rightarrow l'} (\beta),
\end{align}
which implies that one only has to compute the aberration kernel for $m\geq 0$. 
Another compression is caused by the fact that the kernel elements are (steeply) decreasing functions with increasing $\Delta l=l-l'$. Therefore in numerical applications, for every pair $(l', m)$ the kernel elements have to be computed until $l=l'+\Delta l_{\rm max}$, since for $\Delta l>\Delta l_{\rm max}$ the coupling elements vanish for practical purposes. For $\beta \sim 10^{-3}$ one expects $\Delta l_{\rm max}\sim 4-5$ to suffice at $l'\lesssim 3000$.

Finally, according to Eq.~\eqref{eq:Kernel_def_2}, the aberration kernel always acts on spherical harmonic coefficients with fixed $m$. 
It is also clear that $\mathcal{H}^m_{l'\rightarrow l}(\beta)=0$ for $m>l'$, simply because there is no initial combination $(l', m)\rightarrow (l,m)$ with $m>l'$.
This property of the kernel simplifies the computations significantly. 
For example, the monopole can only leak into the $m=0$ terms of the higher multipoles, the dipole can only leak into the $m=\{-1,0,1\}$ terms of the higher multipoles, plus the $m=0$ term of the monopole, the quadrupole can only leak into the  $m=\{-2,-1,0,1,2\}$ terms of the higher multipoles, plus the $m=\{-1,0,1\}$ terms of the dipole, and the $m=0$ term of the monopole, and so on.

\subsection{Recursions for the integrals $\mathcal{H}^m_{l'\rightarrow l} (\beta)$}
\label{app:recursions_kernel}
We now derive simple recursion formulae for the integrals $\mathcal{H}^m_{l'\rightarrow l}(\beta)$ for $l'\leq l$.
It turns out that one also has to restrict the relations derived here to the case $l' > m$. The case $l'\equiv m$ will be discussed below (see Sect.~\ref{app:recursions_Hmm_l}).
We start with the recursion relation
\beal
\label{eq:Plm_recursion}
(l+1-m)\,P^m_{l+1}(\mu')=(2l+1)\,\mu'\,P^m_{l}(\mu')-(l+m)\,P^m_{l-1}(\mu')
\end{align}
for the associated Legendre polynomials. Adding $(2l+1)\,\beta\,P^m_{l}(\mu')$ on both sides of the equation, and multiplying by $P^m_{l'}(\mu) / (1+\beta \mu') $, after integration over $\id \mu'$ one obtains
\beal
\label{eq:rec_HlmII}
\int_{-1}^{1}  P^m_l(\mu')\, \mu\,  P^m_{l'}(\mu)  \id \mu'
=
\frac{(l+1-m)}{2l+1}\mathcal{H}^m_{l'\rightarrow l+1}+\frac{(l+m)}{2l+1}\mathcal{H}^m_{l'\rightarrow l-1}
+\beta \mathcal{H}^m_{l' \rightarrow l}.
\end{align}
The remaining integral can be simplified in the following way:
\beal
\label{eq:rec_Pmu_Int}
\int_{-1}^{1}  P^m_l(\mu')\, \mu\,  P^m_{l'}(\mu)  \id \mu'
&=\int_{-1}^{1}  P^m_l(\mu') \frac{\mu(1+\beta\mu')}{1+\beta\mu'}  P^m_{l'}(\mu)  \id \mu'
\nonumber\\
&=\int_{-1}^{1} \! \frac{P^m_l(\mu') \, \mu\, P^m_{l'}(\mu)}{1+\beta\mu'}  \id \mu' 
+ \beta \int_{-1}^{1}  \frac{P^m_l(\mu') \,\mu'\,\mu \, P^m_{l'}(\mu) }{1+\beta\mu'}   \id \mu'.
\end{align}
Now, using $(2l+1)\,x P^m_l(x)=(l+1-m)P^m_{l+1}(x)+(l+m)P^m_{l-1}(x)$ and rearranging terms, together with Eq.~\eqref{eq:rec_HlmII} we obtain\footnote{For this equation we have assumed that $l' > m$, so that the case $l'\equiv m$ has to be treated separately.}
\beal
\label{eq:rec_Hlm_final}
&\beta \,\mathcal{H}^m_{l'\rightarrow l}
+ \alpha^m_{l}\,\mathcal{H}^m_{l'\rightarrow l-1}
+ \lambda^m_{l} \,\beta\,\mathcal{H}^m_{l'\rightarrow l-2}
\nonumber\\
&\quad\quad
=
\alpha^m_{l'}
\left[
\mathcal{H}^m_{l' -1 \rightarrow l}
+\alpha^m_{l} \, \beta \, \mathcal{H}^m_{l' -1 \rightarrow l-1}
+ \lambda^m_{l} \mathcal{H}^m_{l'-1\rightarrow l-2}
\right]
\nonumber\\
&\quad\quad\quad
-
\lambda^m_{l'}
\left[
\beta\,\mathcal{H}^m_{l'-2\rightarrow l}
+ \alpha^m_{l} \mathcal{H}^m_{l'-2\rightarrow l-1}
+ \lambda^m_{l} \, \beta \, \mathcal{H}^m_{l' -2 \rightarrow l-2}
\right].
\end{align}
Here we introduced the abbreviations $\alpha^m_{l} \equiv [2l-1]/[l-m]$ and $\lambda^m_{l} \equiv \alpha^m_{l}-1$.

The problem is now that numerically the recursion Eq.~\eqref{eq:rec_Hlm_final} is not very stable, since leading order terms cancel each time the recursion is applied.
However, one can use a Taylor series for $\mathcal{H}^m_{l'\rightarrow l}$ and derive recursions for the series coefficients instead.
The parity of the kernel suggests the ansatz 
\beal
\label{eq:H_series}
\mathcal{H}^m_{l'\rightarrow l}=(-1)^{l+l'}\beta^{| l-l' |}\sum_{k=0}^\infty  {}^{(m)}\! \kappa^{l'\rightarrow l}_{2k}\beta^{2k}.
\end{align}
Here we also used the fact, that the kernel coefficients are {\it decaying} with increasing $\Delta l=| l-l' |$. 
Inserting this into Eq.~\eqref{eq:rec_Hlm_final}, and rearranging terms, we obtain
\bsub
\label{eq:rec_kappa_full}
\beal
\label{eq:rec_kappa_full_a}
 {}^{(m)}\! \kappa^{l' \rightarrow l'}_{2k} 
 &=
\alpha^m_{l'}\left[
 \alpha^m_{l'}\,{}^{(m)}\! \kappa^{l'-1 \rightarrow l'-1}_{2k}-2\,{}^{(m)}\! \kappa^{l'-1 \rightarrow l'}_{2k}
 \right]
\nonumber\\[1mm]
&\quad
-
\lambda^m_{l'}\left[
\lambda^m_{l'}\,{}^{(m)}\! \kappa^{l'-2 \rightarrow l'-2}_{2k}-2\,\alpha^m_{l'}\,{}^{(m)}\! \kappa^{l'-2 \rightarrow l'-1}_{2k}
 + 2\, {}^{(m)}\! \kappa^{l'-2 \rightarrow l'}_{2k-2}
 \right]
\end{align}
for $l\equiv l'$. Here ${}^{(m)}\! \kappa^{l' \rightarrow l}_{2k}=0$ for $k<0$.
Similarly, for $l=l'+1$ we have
\beal
\label{eq:rec_kappa_full_b}
 {}^{(m)}\! \kappa^{l' \rightarrow l'+1}_{2k} 
 &=
 \frac{1}{2\alpha^m_{l'+2}}
 \left[
(\alpha^m_{l'+1}\alpha^m_{l'+2}-\lambda^m_{l'+2}) 
 {}^{(m)}\! \kappa^{l' \rightarrow l'}_{2k}
 + 
 \left(\alpha^m_{l'+2}+\frac{\lambda^m_{l'+2}}{\alpha^m_{l'+1}} \right) 
{}^{(m)}\! \kappa^{l' \rightarrow l'+1}_{2k-2}
 -{}^{(m)}\! \kappa^{l' \rightarrow l'+2}_{2k-2}
 \right]
\nonumber\\
&\quad
-\frac{\lambda^m_{l'+1}}{2\alpha^m_{l'+1}}
\left[
\lambda^m_{l'+1}\,
{}^{(m)}\! \kappa^{l'-1 \rightarrow l'-1}_{2k} 
-
2\alpha^m_{l'+1} \,
{}^{(m)}\! \kappa^{l'-1 \rightarrow l'}_{2k} 
+
\frac{\lambda^m_{l'+2} }{\alpha^m_{l'+2} }\,
{}^{(m)}\! \kappa^{l'-1 \rightarrow l'}_{2k-2} 
\right]
\nonumber\\
&\qquad
-\frac{\lambda^m_{l'+1}}{2\alpha^m_{l'+1}\alpha^m_{l'+2}}
\left[
\alpha ^m_{l'+2}\,
{}^{(m)}\! \kappa^{l'-1 \rightarrow l'+1}_{2k-2} 
+
{}^{(m)}\! \kappa^{l'-1 \rightarrow l'+2}_{2k-4} 
\right]
%
%
-\frac{{}^{(m)}\! \kappa^{l'+1 \rightarrow l'+2}_{2k-2}}{2\alpha^m_{l'+1}\alpha^m_{l'+2}},
 \end{align}
and for $l\geq l'+2 $ we find
\beal
\label{eq:rec_kappa_full_c}
{}^{(m)}\! \kappa^{l' \rightarrow l}_{2k} 
&=
\frac{1}{\alpha^m_{l+1}}\,
\left[ \lambda^m_{l+1}\,{}^{(m)}\! \kappa^{l' \rightarrow l-1}_{2k} +{}^{(m)}\! \kappa^{l' \rightarrow l+1}_{2k-2} \right]
\nonumber\\
&\qquad
+\frac{\alpha^m_{l'}}{\alpha^m_{l+1}}\,\left[  \lambda^m_{l+1}  \, {}^{(m)}\! \kappa^{l'-1 \rightarrow l-1}_{2k} 
- \alpha^m_{l+1} {}^{(m)}\! \kappa^{l'-1 \rightarrow l}_{2k-2} 
+ {}^{(m)}\! \kappa^{l'-1 \rightarrow l+1}_{2k-2}  \right]
\nonumber\\
&\qquad\qquad
+\frac{\lambda^m_{l'}}{\alpha^m_{l+1}}
\left[ \lambda^m_{l+1} {}^{(m)}\! \kappa^{l'-2 \rightarrow l-1}_{2k-2} 
- \alpha^m_{l+1}{}^{(m)}\! \kappa^{l'-2 \rightarrow l}_{2k-2}
+{}^{(m)}\! \kappa^{l'-2 \rightarrow l+1}_{2k-4}
\right].
\end{align}
\esub
Equations~\eqref{eq:rec_kappa_full} in principle fully determine the problem and allow us to compute the aberration kernel in a fast way. However, two pieces are missing: (i) the initial condition $\mathcal{H}^m_{m\rightarrow m}$; and (ii) the relations for $\mathcal{H}^m_{m\rightarrow l}$. We derive these in the next two sections.

\subsubsection{Initial condition for the recursions}
\label{sec:initial_cond}
To compute the elements ${}^{(m)}\! \kappa^{m\rightarrow m}_{2k}$ we use the identity $P^m_m(x)=(-1)^m (2m-1)!! (1-x^2)^{m/2}$ and insert this into the definition of $\mathcal{H}^m_{l'\rightarrow l}$. After some algebra, this leads to
\beal
\mathcal{H}^{m}_{m\rightarrow m} (\beta)&=\frac{2^{m+1}}{\gamma^m} 
\sum_{k=0}^\infty  \frac{(2k+m)!}{2^k\,k!}\frac{[(2m-1)!!]^2}{(2m+2k+1)!!}\,\beta^{2k}.
\end{align}
To define the initial conditions for the recursions, it is convenient to redefine the integrals $\mathcal{H}^m_{l'\rightarrow l}\rightarrow\gamma^{-m} \, \mathcal{\tilde{H}}^m_{l'\rightarrow l}$. The recursions Eq~\eqref{eq:rec_kappa_full} remain completely unaltered by this transformation, but one can avoid performing another series expansion of $\gamma^{-m}$ in $\beta$.
Also, since the kernel normalization factors are strong functions of $l$, $l'$ and $m$ (see Eq.~\eqref{eq:Norm_fact}), it is useful to rescale the integrals $\mathcal{\tilde{H}}^m_{l'\rightarrow l}$ in addition, absorbing the factor $\mathcal{N}_{l'm}\,\mathcal{N}_{lm}$. 
Afterwards, the aberration kernel can be expressed as
\bsub
\beal
\label{eq:redef_H}
\mathcal{K}^{l'm}_{lm} (\beta)
&=(-1)^{l+l'}\,\frac{T_0}{T'_0}\,\frac{\beta^{| l-l' |}}{\gamma^{m+1}}\,
\sum_{k=0}^\infty  {}^{(m)} \tilde{\kappa}^{l'\rightarrow l}_{2k}\beta^{2k},
\\[1mm]
{}^{(m)} \tilde{\kappa}^{l'\rightarrow l}_{2k}
&=2\pi\,\mathcal{N}_{l'm}\,\mathcal{N}_{lm}\,{}^{(m)} {\kappa}^{l'\rightarrow l}_{2k},
\\[1mm]
\label{eq:redef_H_c}
{}^{(m)} \tilde{\kappa}^{m\rightarrow m}_{2k}
&=\frac{(2k+m)!}{2^{k}\,k!\,m!}\frac{(2m+1)!!}{(2m+2k+1)!!}.
\end{align}
\esub
After this rescaling, the initial condition for $k=0$ is ${}^{(m)} \tilde{\kappa}^{m\rightarrow m}_{0}=1$. 
It is also clear that for $m\geq 4$ one always finds ${}^{(m)} \tilde{\kappa}^{m\rightarrow m}_{2}>1$. In particular, for $m\gg 1$ and $m\gg k$ one has ${}^{(m)} \tilde{\kappa}^{m\rightarrow m}_{2k} \approx m^k / [(4e)^k\,k!]$, such that for $\beta\sim 10^{-3}$ the series of $\mathcal{K}^{m}_{m\rightarrow m} (\beta)$ always converges rather fast while $m\lesssim 10^4$.
Furthermore, it is directly clear\footnote{This can be also easily shown by expanding $\mathcal{H}^m_{l'\rightarrow l'}$ to lowest order in $\beta$, i.e. {$\mathcal{H}^m_{l'\rightarrow l'}\approx \int P^m_{l'}(x)\,P^m_{l'}(x)\id x = 1$}.} that ${}^{(m)}\! \tilde{\kappa}^{l'\rightarrow l'}_{0}=1$ for all $l'$, since for $\beta = 0$ the aberration kernel should be $\mathcal{K}^{l'm'}_{lm}=\delta_{l'l}\,\delta_{m'm}$.

The rescaling also affects the recursion formulae Eq.~\eqref{eq:rec_kappa_full}. One simply has to multiply all terms with appropriate ratios of $[\mathcal{N}_{l'm}\,\mathcal{N}_{lm}]/[\mathcal{N}_{i m} \mathcal{N}_{j m}]$.
Defining $ \tilde{\alpha}^m_{l}\equiv \sqrt{[4l^2-1]/[l^2-m^2]}$ and $\tilde{\lambda}^m_{l} \equiv\tilde{\alpha}^m_{l}/\tilde{\alpha}^m_{l-1}$, we find that for Eq.~\eqref{eq:rec_kappa_full} one simply has to replace $\alpha \rightarrow \tilde{\alpha}$ and $\lambda \rightarrow \tilde{\lambda}$, without any further changes.

\subsubsection{Recursions for $\mathcal{H}^m_{m\rightarrow l}$}
\label{app:recursions_Hmm_l}
To obtain the required formulae for $\mathcal{H}^m_{m\rightarrow l}$ we start with the identity
\beal
P^{m+1}_{m+1}(\mu)=-(2m+1) \sqrt{1-\mu^2}\,P^{m}_{m}(\mu).
\end{align}
Multiplying on both sides with $P^{m+1}_l(\mu')/[1+\beta\mu']$ and integrating over $\id \mu'$ we find
\beal
\mathcal{H}^{m+1}_{m+1\rightarrow l}&=-(2m+1) \int_{-1}^1 P^{m+1}_l(\mu')\frac{\sqrt{1-\mu^2}}{1-\beta\mu}\,P^{m}_{m}(\mu) \id \mu,
\end{align}
where we made use of the identity $\id \mu' / [1+\beta\mu' ]= \id \mu/[1-\beta\mu] $. Furthermore, 
\beal
\nonumber
\sqrt{1-\mu'^2}=\frac{\sqrt{1-\mu^2}}{\gamma[1-\beta\mu]},
\end{align}
such that we have
\beal
 \int_{-1}^1 P^{m+1}_l(\mu')\frac{\sqrt{1-\mu^2}}{1-\beta\mu}\,P^{m}_{m}(\mu) \id \mu 
 &=   \gamma \int_{-1}^1 P^{m+1}_l(\mu')\sqrt{1-\mu'^2}\,P^{m}_{m}(\mu) \id \mu.
\end{align}
Then with $\sqrt{1-\mu'^2}\,P^{m+1}_l(\mu')=(l-m) \, \mu' P^{m}_l(\mu')-(l+m)\,P^{m}_{l-1}(\mu')$ it is straightforward to show that\footnote{Here the implicit assumption is that $m> 0$.}
\beal
\gamma^{-1}\,\mathcal{H}^{m}_{m \rightarrow l}
&=(2m-1)\left[
(l-1+m)\,\mathcal{H}^{m-1}_{m-1 \rightarrow l-1} + \beta\,(l+1-m) \,\mathcal{H}^{m-1}_{m-1 \rightarrow l} 
\right]
\nonumber\\[1mm]
&\qquad
- \beta\,(l-1+m) \,\mathcal{H}^{m-1}_{m \rightarrow l-1}
- (l+1-m)\,\mathcal{H}^{m-1}_{m \rightarrow l}.
\end{align}
Here again it is better to express this relation in terms of the series coefficients ${}^{(m)}\! \kappa^{l' \rightarrow l}_{2k}$ using the ansatz Eq.~\eqref{eq:H_series}. 
We again replace $\mathcal{H}^{m}_{m \rightarrow l}=\gamma^{-m}\, \mathcal{\tilde{H}}^{m}_{m \rightarrow l}$ and absorb the factor $2\pi\,\mathcal{N}_{l'm}\,\mathcal{N}_{lm}$. 
With this we obtain
\beal
\label{eq:rec_kappa_mm_l}
{}^{(m)} \tilde{\kappa}^{m\rightarrow l}_{2k}
&=
\sqrt{2m+1}\,a
\left[
b\, {}^{(m-1)} \tilde{\kappa}^{m-1\rightarrow l-1}_{2k} -{}^{(m-1)} \tilde{\kappa}^{m-1\rightarrow l}_{2k-2} 
\right]
\nonumber\\
&\qquad\qquad\qquad
+
a \left[
b\, {}^{(m-1)} \tilde{\kappa}^{m\rightarrow l-1}_{2k} -{}^{(m-1)} \tilde{\kappa}^{m\rightarrow l}_{2k} 
\right]
+{}^{(m)} \tilde{\kappa}^{m\rightarrow l}_{2k-2} 
,
\end{align}
where $a\equiv \left[\frac{l-m+1}{2m(l+m)}\right]^{1/2}$ and $b \equiv\left[\frac{2l+1}{2l-1}\,\frac{l+m-1}{l-m+1}\right]^{1/2}$ for $m>0$ and $l>m$.

In a similar way one can show that for $m=0$
\beal
\label{eq:rec_kappa_00_l}
{}^{(0)} \tilde{\kappa}^{0\rightarrow l}_{2k}
&= 
\frac{l}{\sqrt{4l^2-1}}{}^{(0)} \tilde{\kappa}^{0\rightarrow l-1}_{2k} 
+ \frac{l+1}{\sqrt{4l[l+2]+3}}{}^{(0)} \tilde{\kappa}^{0\rightarrow l+1}_{2k-2}.
\end{align}
This equation closes the problem, and we are ready to compute all elements of the aberration kernel to machine precision.
For this one can first generate all the required coefficients $\tilde{\alpha}^m_{l}$ and then compute order by order in $k$ until convergence is reached.
Looking at the properties of the recursion formulae Eq.~\eqref{eq:rec_kappa_full}, \eqref{eq:rec_kappa_mm_l} and \eqref{eq:rec_kappa_00_l}, it seems easiest to use the following procedure:
\begin{itemize}

\item[(i)] Starting with  ${}^{(0)}\! \tilde{\kappa}^{0\rightarrow 0}_{2k}$ for all $k\leq k_{\rm max}$ one can first compute ${}^{(0)}\! \tilde{\kappa}^{0\rightarrow l}_{2k}$ for all required $l>0$ using Eq.~\eqref{eq:rec_kappa_00_l}.

\item[(ii)] Next determine ${}^{(0)}\! \tilde{\kappa}^{1\rightarrow l}_{2k}$ for all required $l\geq 1$, followed by ${}^{(0)}\! \tilde{\kappa}^{2\rightarrow l}_{2k}$, ${}^{(0)}\! \tilde{\kappa}^{3\rightarrow l}_{2k}$ until ${}^{(0)}\! \tilde{\kappa}^{l_{\rm max}\rightarrow l_{\rm max}}_{2k}$, subsequently applying Eq.~\eqref{eq:rec_kappa_full_a}--\eqref{eq:rec_kappa_full_c}.

\item[(iii)]  Using Eq.~\eqref{eq:rec_kappa_mm_l} and the initial condition Eq.~\eqref{eq:redef_H_c} compute all ${}^{(1)}\! \tilde{\kappa}^{1\rightarrow l}_{2k}$ for all required $l>0$.

\item[(iv)] Next determine ${}^{(1)}\! \tilde{\kappa}^{2\rightarrow l}_{2k}$ for all required $l\geq 2$, followed by ${}^{(1)}\! \tilde{\kappa}^{3\rightarrow l}_{2k}$, ${}^{(1)}\! \tilde{\kappa}^{4\rightarrow l}_{2k}$ until ${}^{(1)}\! \tilde{\kappa}^{l_{\rm max}\rightarrow l_{\rm max}}_{2k}$, subsequently applying Eq.~\eqref{eq:rec_kappa_full_a}--\eqref{eq:rec_kappa_full_c}.

\item[(v)] Repeat (iii) and (iv) for all required $2\leq m \leq l_{\rm max}$.

\end{itemize}
\end{appendix}

\end{document}